\documentclass[final]{ws-rv9x6}
\usepackage{caption}
\usepackage{subcaption}
\usepackage{ws-rv-thm}   
\usepackage[square]{ws-rv-van}   
\usepackage[perpage]{footmisc} 
\makeindex

\usepackage{hyperref}
\hypersetup{
   colorlinks,%
   citecolor=blue,%
   filecolor=black,%
   linkcolor=blue,%
   urlcolor=blue
}

\usepackage{amssymb}
\usepackage{pifont}

\newcommand{\cmark}{\ding{51}}%
\newcommand{\xmark}{\ding{55}}%
\newcommand{\dstar}{\ding{72}}%
\newcommand{\dstarr}{\dstar\dstar}%
\newcommand{\dstarrr}{\dstar\dstar\dstar}%

\begin{document}

\chapter[ML scientific competitions and datasets]{Machine Learning scientific competitions and datasets}


\author[D. Rousseau and A. Ustyuzhanin ]{
 David Rousseau\footnote {david.rousseau@ijclab.in2p3.fr} 
  \address{Université Paris-Saclay, CNRS/IN2P3, IJCLab,  91405 Orsay, France}
 Andrey Ustyuzhanin\footnote {andrey.u@gmail.com}
 \address{National Research University Higher School of Economics and Yandex School of Data Analysis, Moscow, Russia}
}


\begin{abstract}
A number of scientific competitions have been organised in the last few years with the objective of discovering innovative techniques to perform typical  High Energy Physics tasks, like event reconstruction, classification and new physics discovery. Four of these competitions are summarised in this chapter, from which guidelines on organising such events are derived. In addition, a choice of competition platforms and available datasets are described.
\end{abstract}

\body

\tableofcontents


\section{Introduction}
\label{s:introduction}
Competitions play an important role in the development of Machine Learning algorithms. The 2012 breakthrough of a Convolutional Neural Network\cite{imagenet_hinton} in the ImageNet Large Scale Visual Recognition Challenge  (ILSVRC)  competition on labelling objects in the ImageNet dataset is often indicated to be the start of the ``Deep Learning revolution". The ImageNet dataset itself is considered as a standard candle in countless papers, as well as for teaching and training. 

Competitions in High Energy Physics are much less part of the culture. One explanation for this is that sharing data has not been the norm, although this is changing. Collaborations of scientists usually analyse the data from the experiment they have built and share the result of the analyses but not the data itself.

For a specific task, e.g. particle identification or event classification, one can find papers on algorithm A applied on dataset alpha; algorithm B applied on dataset beta. The metric used will be similar, at best, identical. However, suppose one sees better result on one side. In that case, it is difficult to infer if algorithm A is intrinsically better than algorithm B or that dataset alpha makes the task easier. If one wants to find a better algorithm, one would gather papers, go to workshops, talk to experts to have suggestions of better algorithms that one would have to re-implement to apply on one's dataset. The difficulty is not just about acquiring the software but also the accompanying expertise.

\label{intro:CTF}Scientific competitions are an alternative approach structured around so-called Common Task Framework (CTF)\cite{donoho50years} that involves:
\begin{enumerate}
    \item A publicly available training dataset involving, for each observation, a list of feature measurements, and a class label for that observation;
    \item A set of enrolled competitors whose common task is to infer a class prediction rule from the training data;
    \item A scoring referee, to which competitors can submit their prediction rule. The referee runs the prediction rule against a testing dataset which is sequestered behind a screen. The referee objectively and automatically reports the score (e.g. prediction accuracy) achieved by the submitted rule.
\end{enumerate}
In reality, the scoring might be quite complicated as translation between domain challenge requirements to a straightforward computational form requires both fluent speaking and understanding potential flaws of both: the domain and machine learning languages.  

The objective of this chapter is to look under the hood of scientific competitions and encourage participation and foster the organisation of future competitions. 

The chapter is organised as follows. Section 2 to 5 give a summary of four physics competitions and related datasets. Section 2 HiggsML on event classification, section 3 Flavour of Physics on event classification in the presence of mismodelings, section 4 TrackML on pattern recognition, section 5 LHC-Olympics on anomaly detection. Section 6 lists available competition platforms and section 7  lists available open datasets. Section 8 indicates general guidelines for scientific competition organizers, based on the experience organising such competitions. Section 9 is the conclusion.


\section{HiggsML}
\label{s:higgsml}
The Higgs Boson Machine Learning challenge (HiggsML in short)\footnote{https://higgsml.lal.in2p3.fr} took place on the Kaggle platform in 2014. At the time, Machine Learning was already used at the LHC experiments (see \cite{Radovic:2018dip} although some of the results quoted there are post-2014) but in most cases, this was Boosted Decision Trees, while the Deep Learning revolution had already started elsewhere. Some breakthrough papers (in particular \cite{Baldi:2014kfa} 
) were indicating a potential for Deep Learning for final analysis. 
The motivation for the HiggsML challenge was to reach out to the Computer Science community to explore the possibilities of modern Machine Learning algorithms for a classification problem pertaining to Higgs boson physics at the LHC.
The HiggsML challenge is described in details in \cite{higgsml_cern_odp_document}, which is the document accompanying the final release of the dataset\cite{higgsml_cern_odp_dataset} on the CERN Open Data Portal; lessons derived from the challenge are described in \cite{cowa14} with more details in the write-up of contributions of the dedicated HEPML workshop which has taken place at NeurIPS~2014\cite{HEPML2014}.  Only a summary is given here.

\subsection {Dataset and score}
\label{sec:higgsml_dataset_score}
A dataset of 250.000 events (out of an original dataset of about 800k events, the complement being held out for evaluation) was provided, each event providing 30 features, measurements from simulation proton collision, which had been used by the ATLAS experiment for its first paper on the specific topic \cite{Aad:2015vsa}. The events were either from Higgs boson decay into a tau-lepton pair (the signal) or from other processes, Z boson decaying into a tau-lepton pair as well, top pairs and W.
There were two classes of features, the primary and the derived ones. The primary ones were essentially the 3-momentum of key event particles: an electron or muon, a $\tau$ hadron decay, the missing transverse energy which is a 2D pseudo particle, and the possible leading and subleading jets.   
The derived parameters are features (which could be recomputed from the primary ones) defined in the same ATLAS paper \cite{Aad:2015vsa}, that offer a good separation between signal and background. These derived features had been crafted by physicists to maximise the separation between signal and background.
It should be noted that not all events have two jets (some have zero, some have one) so that the jet quantities (primary or derived) might be absent for some events.
For training, in addition to the label ``S" for signal or ``B" for background, a weight is also given, which allows computing the expected number of signal and background events (for 2012 LHC luminosity).  

A non-standard (for Machine Learning) figure-of-merit was used to rank the criterion, the Approximate Median Significance (AMS), which quantify (in number of standard deviations) the expected discovery significance of an experiment. It is obtained with the following formula where $s$ (resp. $b$)is the number of expected signal (resp. background) events : 
\begin{equation}\label{eq:higgsml:score}
AMS = \sqrt{2} \sqrt {(s+b+10) \log (1+s/(b+10))-s}
\end{equation}
, where $s$ (resp. $b$) is the sum of the weights of signal (resp. background) events passing the selection in the test sample.
This is the usual (for physicists) so-called Asimov formula~\cite{Cowan:2010js}, except that 10 is added to $b$ as a regularisation term, to avoid nonphysically large significance in the very strong selection regime, where $b$ can be less than 1. The impact of using this figure of merit compared to e.g. a more classical ROC-AUC or accuracy criteria is that more importance is given to the part of the ROC curve with large background rejection and small signal efficiency (large True Negative and small True Positive), see Sec.\ref{sec:FOP_eval_FOM} for a different means to achieve the same goal.

\subsection {Competition}

\begin{figure}[ht]
\centering

\includegraphics[width=.6\textwidth]{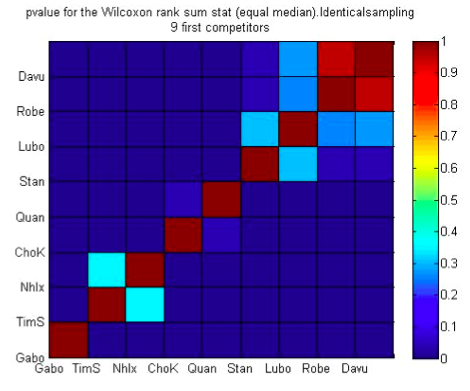}
\caption{(from \cite{cowa14}) p-values of the pairwise Wilcoxon rank sum test}
    \label{fig:higgsml_wilcoxon}       
\end{figure}

\begin{figure}[ht]
\centering

\includegraphics[width=\textwidth]{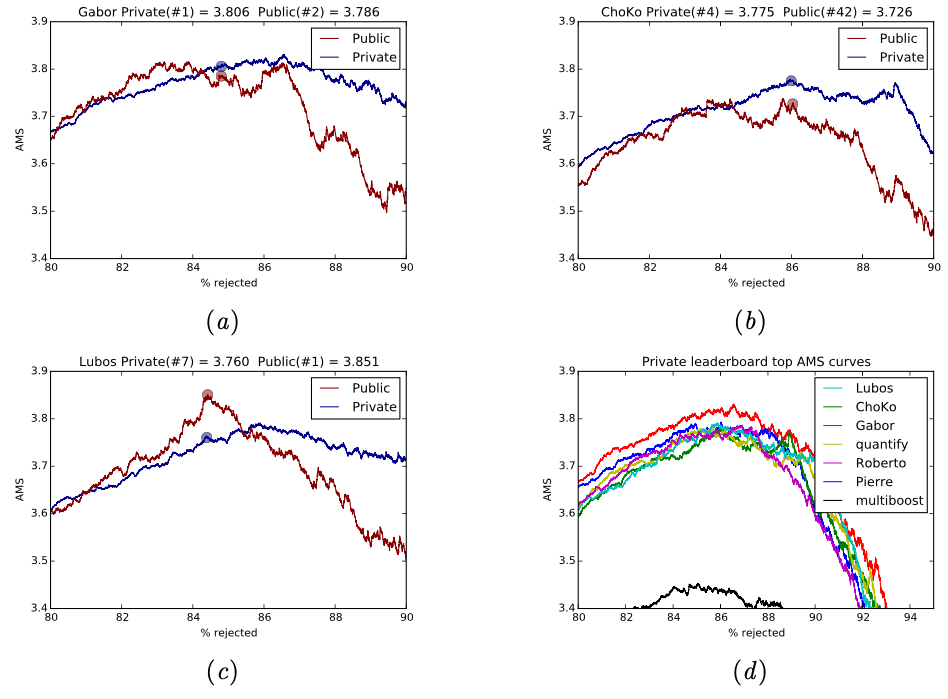}
\caption{(from \cite{cowa14}) AMS curves for some participants  comparing performance between the public and private test sample (a,b,c) and for several participants on the private test sample (d). The horizontal axis is the weighted proportion of selected background events.} 
    \label{fig:higgsml_overtraining}      
\end{figure}

The participation was large with close to 2000 participants, a record at the time for Kaggle challenges.
The ranking among the top 10 was tight, with some rank changes when the private leaderboard (established on a preserved dataset) was revealed. Subsequent studies with a bootstrap technique showed that number 1 (\texttt{Gabor Melis}) rank was solid, while number 2 (\texttt{Tim Salimans}) and number 3 (\texttt{nhlx5haze}) could have exchanged places, but where well separated from number 4 and beyond (see Fig.~\ref{fig:higgsml_wilcoxon}). Fig~\ref{fig:higgsml_overtraining} shows the AMS performance for some top participants. Particularly  interesting curves are the ones from \texttt{Lubos Motl's Team} who was number 1 on the public leaderboard but fell to number 8 on the final leaderboard. A sharp peak on the public test curve (with no counterpart on the private test curve) is due to public leaderboard overfitting as the team has claimed to ``play" the public leaderboard, adjusting parameters in a semi-automatic fashion to improve their public score.

One key feature provided was DER\_mass\_MMC, which is an estimator of the mass of the $\tau^+\tau^-$ pair obtained through a complex MCMC estimation. Although all the inputs to do this calculation were available, the software to compute it has not been released. Nevertheless, \texttt{kesterlester}, a physicist, has released the result of a similar computation under name \texttt{Cake}\footnote{\url{https://www.kaggle.com/c/higgs-boson/discussion/10329}}. It improved significantly the results of most participants, but the top ones did not see any improvements using it, most likely because their classifiers were already able to extract sufficient information from the features they had built. 

The number 1 \texttt{Gabor Melis} used an ensemble of dense neural network with 3~hidden layers; however, from his assessment, he got an edge through careful use of nested Cross-Validation. Fig~\ref{fig:higgsml_overtraining}.a shows his AMS performance on the public test set to be relatively flat compared to others, while Fig~\ref{fig:higgsml_overtraining}.d shows it is clearly above the others almost everywhere. Number 2 \texttt{Tim Salimans} has used an ensemble of Regularized Gready Forest and number 3 \texttt{nhlx5haze} an ensemble of neural networks. Tianqi Chen and Bing Xu (team \texttt{crowwork}) have reached a modest rank 45, however, they got the special Hep ML prize from the jury, for they have released early in the competition their new Boosted Decision Tree algorithm XGboost\cite{Chen:2016:XST:2939672.2939785} and supported its  use. XGboost was used by many participants including top 10.
XGboost popularity has grown ever since and has been one of the primary tools used in Kaggle competitions\cite{xgboost_kaggle_chollet}. Its usage is also growing in high energy physics publications (e.g.~\cite{Aaboud:2017jvq}), given its high speed and classification performance.

\subsection{Follow-up studies}

The complete HiggsML dataset has been released on the CERN Open Data portal \cite{higgsml_cern_odp_dataset}. The full 818,238  events dataset has been released.  The team has hesitated to release it completely, without holding a reserve test set. The benefit is that future users have the largest statistics (compared to the 250,000 Kaggle participants had). The downside is that there is no possibility for independent check for overfitting.

The HiggsML dataset (either from Kaggle or from CERN ODP) has been used extensively since the competition for various tutorials (e.g. \cite{higgsml_tutorial_hsf}), courses (e.g. \cite{higgsml_course_schwander}), blog posts (e.g. \cite{higgsml_blog_wittenauer}), PhD dissertations (e.g. \cite{higgsml_phd_chakravarti} )  and papers (see later in this section). While most are informative, some common mistakes have been seen, especially concerning the weight: 
\begin{itemize}
    \item as the weights are such that events are normalised to 2012 Large Hadron Collider data taking, an event weight reflects the way it was generated so that the weight is an almost certain give away of the signal or background label of the event. For this reason, during the Kaggle competition, it was only provided for the training sample, not the test sample. However, in some follow-up studies,  some people have used the weight as a regular feature (although being strongly advised against it through the accompanying document) which give them extremely good performances. Since many people do not read the documentation, probably the weight should have been renamed \texttt{weight\_DO\_NOT\_USE\_IN\_TRAINING}
  \item in contrast with AUC or accuracy, the AMS does depend on the total weight of each sampling (in a trivial way AMS$\simeq \frac{s}{\sqrt{b}}$, so simply using half of the dataset divides AMS by $\sqrt{2}$). The dataset documentation does specify that whenever a subsample of the dataset is used for evaluation, weights should be scaled up by the inverse of the fractional size of the subsample (so a factor 2 for a subsample of $\frac{1}{2}$). This was not done correctly in some follow-up studies. 
\end{itemize}
Besides, claims for AMS well above 3.81 reached by the winner of the competition are most likely due to overtraining. 

A recent thorough post-challenge analysis was done \cite{Chatham_Strong_2020}, where the author has studied data augmentation, learning rate and momentum scheduling, (advanced) ensembling in both model-space and weight-space, and alternative architectures and connection methods, using a modern NN library. He reaches the same 3.81 AMS although with considerably faster training time.

Beyond classification, a python script allows introducing systematic effects (miscalibration or poorly known backgrounds) \cite{victor_estrade_2018_1887847} and has been used to investigate how to deal with systematic effects \cite{Estrade:DLPS2017,estradephd,Wunsch_2020,Wunsch:2020iuh,Dorigo:2020ldg}.

\section{Flavour of physics}
\label{s:fop}

\subsection{Intro}
Offline data analysis in particle physics has many challenges that can provoke communications between the physics and data science communities. In addition to sensitivity increase, there are questions of a) training ML algorithms using the mixture of real and simulated samples (see Sec.~\ref{sec:FOP_eval_agreement}) and b) reducing the impact of so-called \textit{nuisance parameters} that affect the likelihood and posterior distributions non-systematically (see Sec~\ref{sec:FOP_eval_correlation}). LHCb collaboration\footnote{main contributors: Thomas Blake, Marc-Olivier Bettler, Marcin Chrzaszcz, Francesco Dettori, Andrey Ustyuzhanin and Tatiana Likhomanenko} prepared a competition to address those challenges via a competition on Kaggle platform \footnote{\url{https://www.kaggle.com/c/flavours-of-physics/}} that was active for three months in 2015. 

\subsubsection{The challenge goal}
The main goal of this challenge is to gain sensitivity in the search for $\tau^{-}\rightarrow\mu^{-}\mu^{-}\mu^{+}$ decays. That is achieved by improving the discriminating power between signal events (where the decay did occur) and background events (where it did not). LHCb collaboration provides signal and background samples for training and testing. The evaluation happens in three steps: firstly, the classifier is checked not to depend too strongly on the discrepancies between real data and simulation. Then it checks if the classifier output is decorrelated with the $\tau$ mass. Finally, the comparison of the classifiers is performed using the weighted area under the ROC curve.

\subsubsection{Physics motivation} 
The search for decays that do not conserve primary particle flavour quantities started in the late 1930s with the discovery of the muon ($\mu$). It was believed that muons were an excited electron state in which case one would expect to observe a decay $\mu^{-} \to e^{-}\gamma$, where a photon with predictable energy would be emitted. No such process has ever been observed. The muon decays instead through the process $\mu^{-} \to e^{-}\nu_{\mu}\bar{\nu}_{e}$, with the emission of a muon neutrino and an electron anti-neutrino to preserve the total electronic and muonic lepton numbers.
Similarly, in the 1970s an even heavier lepton was discovered as product of $e^{+}e^{-}$ annihilation: the tau ($\tau$) lepton, with a mass equivalent to about 3500 electrons. Typical decays of the $\tau$ leptons are $\tau^{-} \rightarrow e^{-} \nu_{\tau}\bar{\nu}_{e}$ and $\tau^{-} \rightarrow \mu^{-} \nu_{\tau} \bar{\nu}_{\mu}$, that conserve the various lepton numbers involved.
However, if lepton flavour is not a perfectly conserved quantity in nature, and various explanations of the matter asymmetry in the universe require this, then the $\tau$ lepton can also decay into three muons though the reaction $\tau^{-}~\rightarrow~\mu^{-}~\mu^{-}~\mu^{+}$, forbidden instead in the Standard Model. 

The discovery of such a reaction would therefore be a breakthrough on the laws of nature and a sign of long-sought new physics. Search for those decays performed by LHCb collaboration at that time is described in\cite{aaij2015search}.

\subsection{Data description}
In this competition, participants were given a list of collision events and their properties. They had to predict whether a $\tau \to 3 \mu$ decay happened in this collision. This $\tau \to 3 \mu$ is currently assumed by scientists not to occur, and the goal of this competition is to discover $\tau \to 3 \mu$ happening more frequently than scientists now can understand.

\subsubsection{Signal channel}
Different mechanisms can produce tau leptons (mass of $\tau$ equals to 1776.86~$\mathrm{MeV}/c^{2}$). At LHCb, taus are produced in the decay of heavy flavoured particles (containing a \textit{c} or a \textit{b} quark), which are listed in Table~\ref{tab:signal}.
They are mainly produced in the decays of $D^{-}_{s}$ or
$D^{-}$ particles, such as $D^{-} \to \tau \eta$. In the simulation samples provided, the correct proportions of the different tau production mechanisms are respected, and the feature \texttt{production} identifies the production mechanism.

\begin{table}
\tbl{Production mechanisms and their proportions for tau leptons at LHCb, according to the centre-of-mass energy. $X_b$ denotes any particle containing a beauty (\textit{b}) quark. \texttt{production} is a label that, denotes the production mechanism of $\tau$ for simulated decays. In the data, this label is set to -99.}{%
\begin{tabular}{cccc}
\hline Mode & $7 \mathrm{TeV}$ & $8 \mathrm{TeV}$ & production \\
\hline Prompt $D_{s}^{-} \rightarrow \tau$ & $71.1 \pm 3.0 \%$ & $72.4 \pm 2.7 \%$ & 1 \\
Prompt $D^{-} \rightarrow \tau$ & $4.1 \pm 0.8 \%$ & $4.2 \pm 0.7 \%$ & 2 \\
Non-prompt $D_{s}^{-} \rightarrow \tau$ & $9.0 \pm 2.0 \%$ & $8.5 \pm 1.7 \%$ & 5 \\
Non-prompt $D^{-} \rightarrow \tau$ & $0.18 \pm 0.04 \%$ & $0.17 \pm 0.04 \%$ & 6 \\
$X_{b} \rightarrow \tau$ & $15.5 \pm 2.7 \%$ & $14.7 \pm 2.3 \%$ & 4 \\
\hline
\end{tabular}}
\label{tab:signal}
\end{table}

\subsubsection{Background}
\begin{figure*}
\begin{center}
\includegraphics [width=\linewidth]{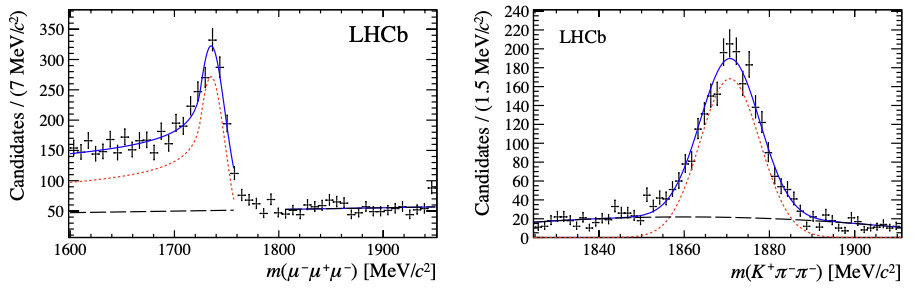}
\caption{\label{fig:background}D meson invariant mass distribution in $D^{+} \rightarrow K^{-} \pi^{+} \pi^{+}$ decays as observed in data. On the left-hand side, all hadrons have been assigned muon mass hypothesis before computing the mass of the mother particle, on the right-hand side, the correct mass hypotheses have been used. }
\end{center}
\end{figure*}

The background for $\tau^{-}\rightarrow \mu^{-}\mu^{-}\mu^{+}$ decay can be divided into two categories. The first one consists of decays with one or more light hadrons (pion or kaon) is wrongly identified as a muon. The main process in this category is $D^{+} \rightarrow K^{-} \pi^{+} \pi^{+} .$ The invariant mass distributions for this process are shown in Fig.\ref{fig:background}. These two mass distributions differ because of the mass assigned to the final states and thus used when computing the mass of the initial state. On the left-hand side, the muon mass $\left(105.66 \mathrm{MeV} / c^{2}\right)$ is assigned to all final states, while, on the right-hand side, the correct masses for kaons and pions $\left(139.57 \mathrm{MeV} / c^{2}\right.$ for $\pi^{\pm}$ and 493.68 for $\left.K^{\pm}\right)$ are used. Hence the misidentification results in a shift in the mass of the initial state.

The second dangerous background $D_{s} \rightarrow \eta\left(\rightarrow \mu^{+} \mu^{-} \gamma\right) \mu^{-} \nu_{\mu}$ originates from the decay in which there are three real muons that can mimic the signal signature. This background can be effectively removed requiring all mass combinations of two muons of opposite sign to be greater than 450~$\mathrm{MeV} / c^{2}$.

\subsubsection{Additional data}
Training of a classifier that is capable of discriminating the signal from the background is a delicate procedure since one can induce unwanted systematic biases that would affect the physics estimations in unpredictable ways. Additional datasets described in the following subsection were supplied to mitigate the risk.

\subsection{Evaluation procedure}
\subsubsection{Verification of the agreement}
\label{sec:FOP_eval_agreement}
Participants have trained classifier models on simulation data for the signal and real data side-bands for the background, so it is possible to reach a high performance by picking features that are not perfectly modelled in the simulation (\texttt{min\_ANNmuon} is an example of such feature). Organizers demand the classifiers not to have large discrepancy when applied to data and simulation. To estimate the discrepancy a \textit{control channel}, 
$D_{s}^{+} \rightarrow \phi\left(\rightarrow \mu^{-} \mu^{+}\right) \pi^{+},$ is used. It has a similar topology as the signal decay. 
Organizers provide both data and simulation samples for this decay in the \texttt{check\_agreement.csv} data set. The Kolmogorov-Smirnov (KS) test is used to evaluate the differences between the classifier distributions on both data sets. The evaluation constraint was that the KS-value on the test data set has to be smaller than~0.09. 

The cumulative distribution (CDF) functions are computed for simulated data predictions, and real data predictions and Kolmogorov-Smirnov metric is calculated:
$$
K S=\max \left|F_{\text {simulation }}-F_{\text {real }}\right|
$$
where $F_{\text {simulation }}$ and $F_{\text {real }}$ are cumulative distribution functions for Monte Carlo data and real data correspondingly.

\subsubsection{Verification of the correlation}
\label{sec:FOP_eval_correlation}
Correlation of classifier output with the $\tau$ mass are unfavourable for data analysis, since those correlations can cause an artificial signal-like mass peak or lead to incorrect background estimations. To prevent cheating of organizers have introduced the Cramer-von Mises (CvM) test\cite{cramr1928composition} to estimate the degree of correlation between the prediction and the mass. The CvM-value of the test has to be smaller than~0.002. Organizers have included the script for computing the CvM-value, so participants could verify own models on \texttt{check\_correlation.csv}. 
CvM metric for the whole dataset predictions CDF is compared to a local (in some mass range) predictions CDF. After that, all intervals are averaged:

\begin{equation}
\begin{aligned}
C v M_{\text {interval }} &=\int\left(F_{\text {global }}-F_{\text {interval }}\right)^{2} d F_{\text {global }} \\
C v M &=<C v M_{\text {interval }}>_{\text {interval }}
\end{aligned}
\end{equation}

where $F_{\text {global }}$ and $F_{\text {interval}}$ are predictions cumulative distribution functions for all the data and data in a given local mass range.

\subsubsection{Figure of merit}
\label{sec:FOP_eval_FOM}
The calculation of the final figure of merit is performed only if the two above tests are passed on \texttt{test.csv} with success and is calculated only using events with \texttt{min\_ANNmuon} $\geqslant 0.04$.

\begin{figure*}
\begin{center}
\includegraphics [width=0.7\linewidth]{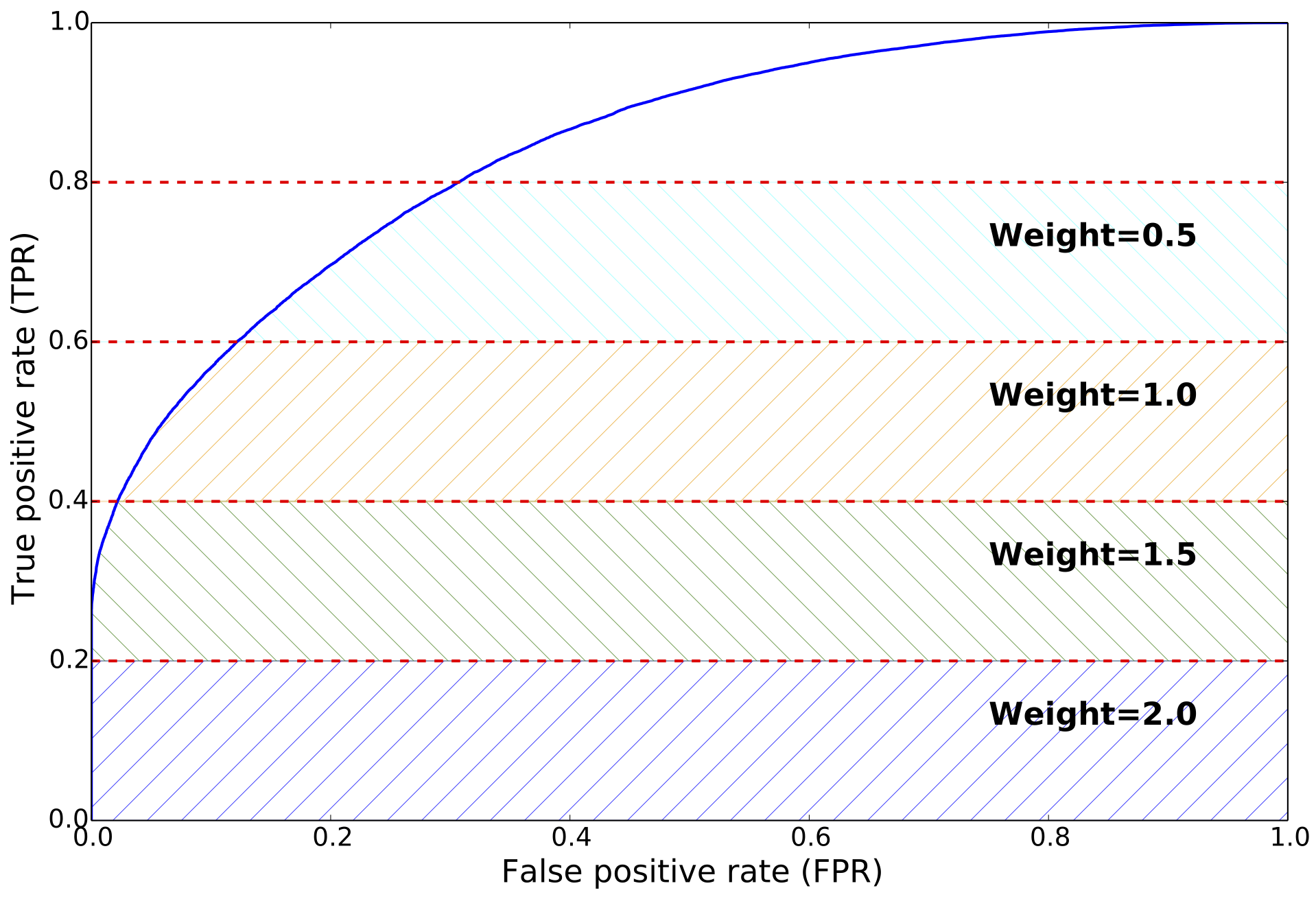}
\caption{\label{fig:weighted_roc}Weights assigned to the different segments of the ROC curve for the purpose of submission evaluation. The \textit{x} axis is the False Positive Rate (FPR), while the \textit{y} axis is True Positive Rate (TPR).}
\end{center}
\end{figure*}

Initially, the LHCb used the CLS
\cite{Read:2002hq} method to determine the upper limit. This method, unfortunately, is computationally expensive and can't be used in this challenge. Instead, organizers proposed a much simpler metric, which is the weighted area under the ROC curve. The regions and their weights are illustrated by Fig.\ref{fig:weighted_roc}.

The reason to assign different weights to different bins of signal efficiency is that the sensitivity to a given process is not a linear function of expected background events. Most of the sensitivity is obtained when the number of expected background events is $\mathcal{O}(1).$ (see Sec.\ref{sec:higgsml_dataset_score} for a different means to achieve the same goal). For example see Table 208 in Ref.\cite{amhis2014heavy}.

\subsubsection{Competition datasets}
All the competition data is provided in the following files:
\begin{enumerate}
    \item \texttt{training.csv} is a labelled data set (the \texttt{signal} being 1 for signal events, 0 for background events) to use for training the classifier. Background events come from real data mass side-bands and from the simulation.
    \item \texttt{check\_agreement.csv} is a labelled data set (the \texttt{signal} being 1 for simulated data, 0 for real data) with the same features as in the \texttt{training.csv}. This data set is used to check the agreement between simulated and real data as described in Sec.\ref{sec:FOP_eval_agreement}.
    \item \texttt{check\_correlation.csv} is a data set with the same features as the \texttt{training.csv}, to check correlation of the classifier with the $\tau$ mass as described in Sec.~\ref{sec:FOP_eval_correlation} before submission.
    \item \texttt{test.csv} is a non-labelled (signal and background are mixed) data set, containing a)~simulated signal events and real background data, b)~simulated events and real data for the control channel.
\end{enumerate}

The setup for the challenge was unusually complicated. The correlation and agreement checks was introduced specifically to match the intuition of physics checks with Kaggle platform requirements. So the organizers have prepared special kind of prizes to the community to mitigate the risk that smart participants find a way to bypass those checks.

\subsection{Prizes and participation statistics}
\begin{figure*}
\begin{center}
\includegraphics [width=0.8\linewidth]{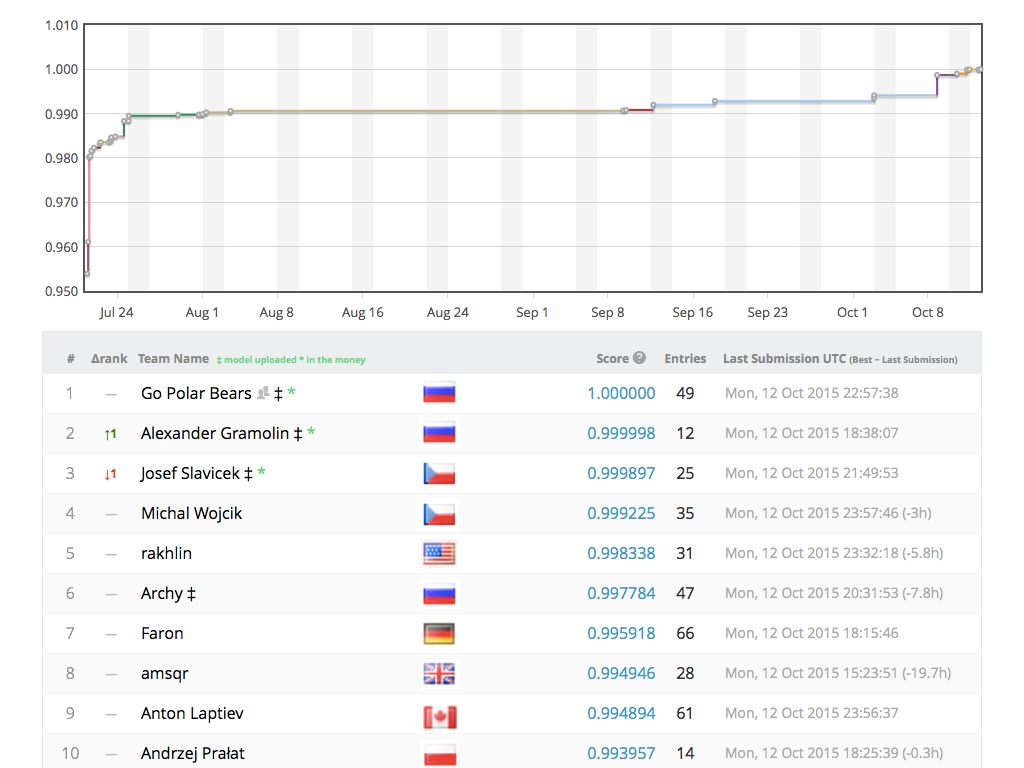}
\caption{\label{fig:FOP-private_lb}Top private `Flavour of Physics` leaderboard score evolution.}
\end{center}
\end{figure*}

The competition was running for three months and has attracted 673 teams. Participants submitted more than 10 thousand different solutions. The main prize allocation for the competitors was: USD 15000 as judged by official Kaggle leaderboard. Additionally, a special Physics Prize that was awarded to teams that, as judged by the LHCb collaboration members, create a model that is most useful from the physical perspective. The main motivation for the special prize was that the challenge setup was quite tricky for regular Kaggle challenge. So it could provide an incentive for the participants to find workarounds to bypass the checks and produce meaningless solutions from the Physics perspective. The Fig.\ref{fig:FOP-private_lb} shows the private leaderboard statistics. One can see that the top teams have reached the ideal score of 1.0, which turned out to be a clever way to bypass additional checks while still introducing unwanted selection properties. Indeed, the trick of exponentiation\footnote{\url{https://www.kaggle.com/rakhlin/abcde/code}} allowed to get a rather high score in the leaderboard. Organizers have scrutinized solutions by top 20 participants and have identified several solutions that avoided unphysical score overfitting (see Section \ref{sec:FOP_physics_prize}).

\subsection{Physics prize and follow-up workshop}
\label{sec:FOP_physics_prize}
The Heavy Flavour Data Mining workshop was organized at the University of Zurich in February 2016\footnote{https://indico.cern.ch/event/433556/} to wrap-up the results of the competition and to award the physics prizes. The recipients were Vicens Gaitan and Alexander Rakhlin. The main ideas of their approaches are highlighted below.
\subsubsection{Data doping by Vicens Gaitan}
\begin{figure*}
\begin{center}
\includegraphics [width=0.8\linewidth]{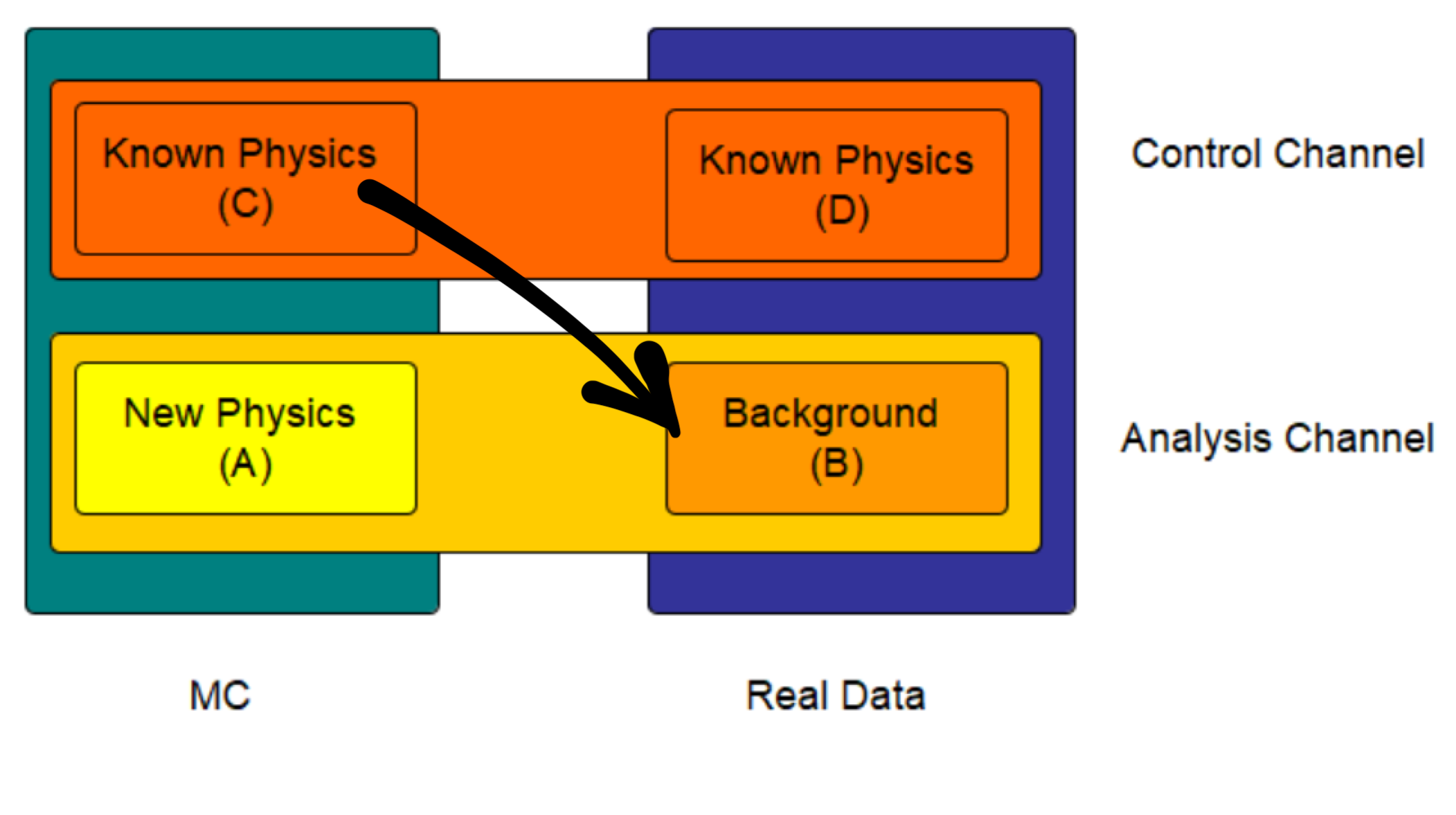}
\caption{\label{fig:FOP-doping}Data doping: the training set with a small number of Monte Carlo events from the control channel, but labeled as background.}
\end{center}
\end{figure*}

The idea is to ``dope" (in the semiconductor meaning) the training set with a small number of Monte Carlo events from the control channel but labelled as background. It disallows the classifier to pick features discriminating data and Monte Carlo. Fig.\ref{fig:FOP-doping} illustrates the doping procedure.

Such a procedure involves two parameters that regularize the learning: a) The number of ``doping" events and b) the complexity of the classifier (for instance number of trees). Those can be tuned depending on the problem and data at hand.

\subsubsection{Transfer learning by Alexander Rakhlin}
The network is trained in the two-stage process. 
\begin{enumerate}
    \item create a strong model for the signal channel using all available features. This model is based on an ensemble of 20 feed-forward fully connected neural nets. 
    \item transfer this model to control channel. The model's output is stacked with original features and cascaded to additional \textit{``transductive"} neural network of similar configuration. The purpose of the second net is to track the first model's output on the signal channel with minimal and controlled adaptation to control channel.
\end{enumerate}
All three metrics (AUC, KS, CVM) are global and not analytically differentiable. It makes gradient descent generally impossible. So the procedure is the following. Initial weights of the transductive network are set to reproduce the output of the original model; this is accomplished after 1-3 epochs of standard GD with cross-entropy loss on the signal channel. Adaptation of weights is made with stochastic optimizer using Powell's method. Loss function incorporates AUC, KS, CVM metrics and allows controlling them during optimization. As a result, it obtains best of the two worlds: performance on signal channel preserved as much as possible (slightly drops only in the 3rd decimal place), the tests on control channel passed. To keep the model as physically sound, one can control its performance on the signal channel during optimization. Furthermore, it is possible to restrict the transductive network from excessive deviation from its original state (weights) or implement any other regularizer\footnote{\url{https://github.com/alexander-rakhlin/flavours-of-physics}}. 

\subsection{Conclusion}
The Flavours of Physics challenge was aiming at an ambitious goal of finding a way to deal with nuisance parameters and MC/real data discrepancies happening in particle physics analyses. In real-life settings, all the checks are performed by professionals with physical intuition that helps them to keep solutions under meaningful constraints. Translation of those constraints to the competition platform implies taking the risks of a) simplifying of the limitations and b) giving incentive to the participants to hack the metric. It has happened to the challenge, and one can follow the details on Kaggle forum. Nevertheless, special prizes allocated by the organizers, which had to be awarded by the domain scientist committee has allowed motivating development of physically-sound solutions that were outlined above.

\section{TrackML}
\label{s:trackml}

\subsection {Introduction}

The Tracking Machine Learning (TrackML) challenge\footnote{\url{https://sites.google.com/site/trackmlparticle/}} took place mainly in two phases, an Accuracy phase\cite{TrackMLAccuracy2019} in 2018 on the Kaggle platform\footnote{https://www.kaggle.com}, and a Throughput phase \cite{TrackMLThroughput2020} (combining accuracy and speed) in 2018-2019 on Codalab\footnote{https://competitions.codalab.org}, preceded by a limited scope 2D prototype competition in April 2017 \cite{TrackMLRamp2017}. The challenge is described in details in the papers referenced above, only a summary is given here, focusing more on the lessons (see also \cite{Rousseau2019}).
 
\begin{figure*}
\begin{center}
\includegraphics [width=\linewidth]{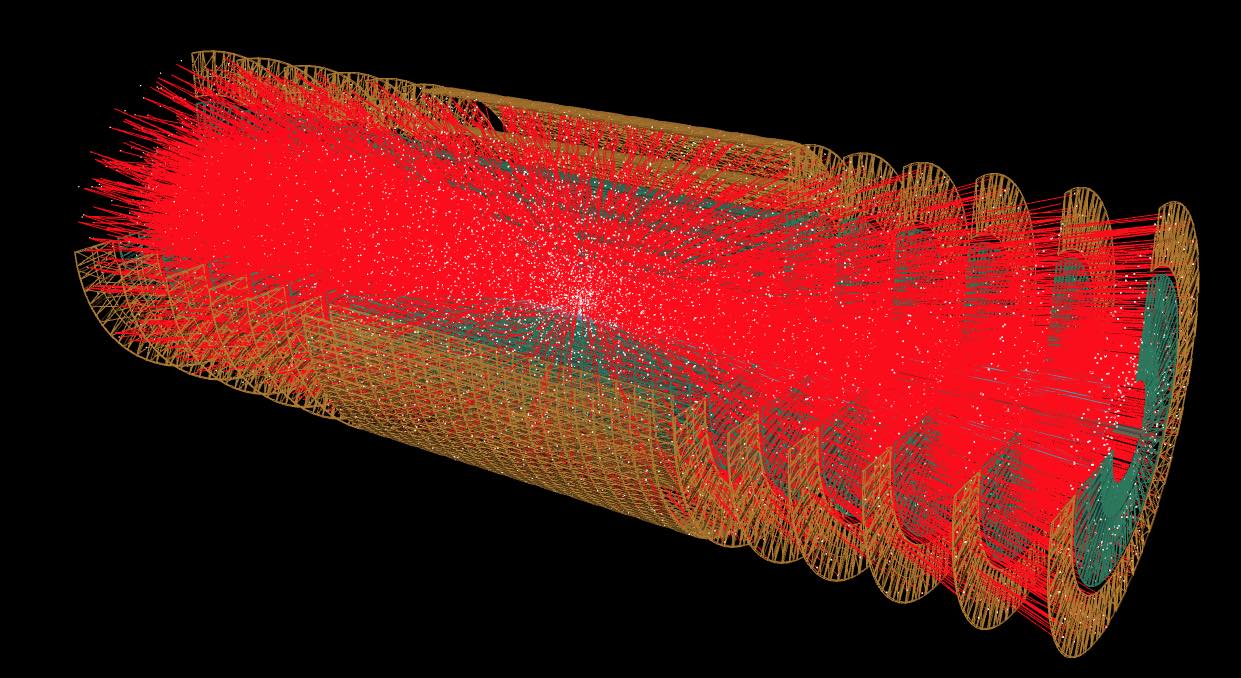}

\caption{(from\cite{TrackMLAccuracy2019}) TrackML detector (one sector of the detector has been etched out). White dots are the measured points, while the red lines are the trajectories of the particles.}
\label{fig:3d_display}
\end{center}
\end{figure*}

The analysis pipelines of the proton collisions at the Large Hadron Collider (or {\it events}) rely on a first step, the reconstruction of the trajectories of the particles within the innermost parts of the detector. 
The time to reconstruct the trajectories --- in a constant magnetic field these would follow a helical path --- from the measurements (3D points) is expected to increase faster than the projected computing resources. New approaches to pattern recognition are thus searched for to exploit fully the discovery potential of the High Luminosity-Large Hadron Collider. 
A typical event for ATLAS or CMS detector at HL-LHC design luminosity would have about 100,000 points to be associated into 10,000 trajectories (see Fig.\ref{fig:3d_display}). The state of the art was about 10s per event on a modern CPU when the challenge was designed. Given that 10 to 100 billions such collisions need to be treated each year the importance of a significant increase of the reconstruction throughput becomes evident.

The goal of the TrackML challenge was to expose the problem of fast particle tracking to the wide Computing Science community. Since designing new algorithms and writing fast software require separate expertise, it was decided early to split the competition into two phases: the first phase (Accuracy) would only be about algorithm accuracy, while the second (Throughput) would be about fast software with a good compromise on accuracy. 

While for the Accuracy phase, participants had to upload a solution file to Kaggle platform indicating how the points are clustered into tracks, for the Throughput phase participants had to upload their software directly to the Codalab platform, on which it was executed in a controlled environment. By doing so, the resource usage was measured in a standardized way, and the Throughput score was then derived from the accuracy and the speed.

\subsection {Dataset and score}

A dataset consisting of an accurate simulation\cite{ACTS} of an LHC-like experiment has been created, listing for each event the measured 3D points, and the list of 3D points associated to a true track. 

The detector simulated is a full Silicon detector organised in cylinders and disks sharing the same axis of symmetry $z$; the origin of the axis is at the centre of symmetry. An approximately solenoidal magnetic field of the same axis bends the particles so that their trajectories are approximate arc of helices. Most, but not all particles, start from close to the origin.

The participants to the challenge should find the tracks, meaning building the list of 3D points belonging to each track, in an additional test dataset without the ground truth. 

Detailed algorithm performance studies usually involve in-depth analysis of hundreds of histograms. Yet, as usual for a competition, algorithms should be ranked from a single score number to be maximised. The Accuracy score was chosen to be  ``the weighted fraction of points correctly assigned", which is computed from the point association inferred by the participants. This choice for an Accuracy score based on points was somewhat counter-intuitive, as it is much more usual in the community to evaluate the performances in term of found tracks, examining the tracking efficiency (fraction of tracks found) and quality (precision of the reconstructed track parameters). 
The post-competition in-depth analysis of the algorithms submitted has shown that this choice has been correct, as the best algorithms in terms of the score also had the best performances when analysed in depth.
It should be noted that this score is non-standard and required specific development by the Kaggle Data Scientist in charge of the competition.

\begin{figure}[ht]
\centering
\includegraphics[width=.8\linewidth]{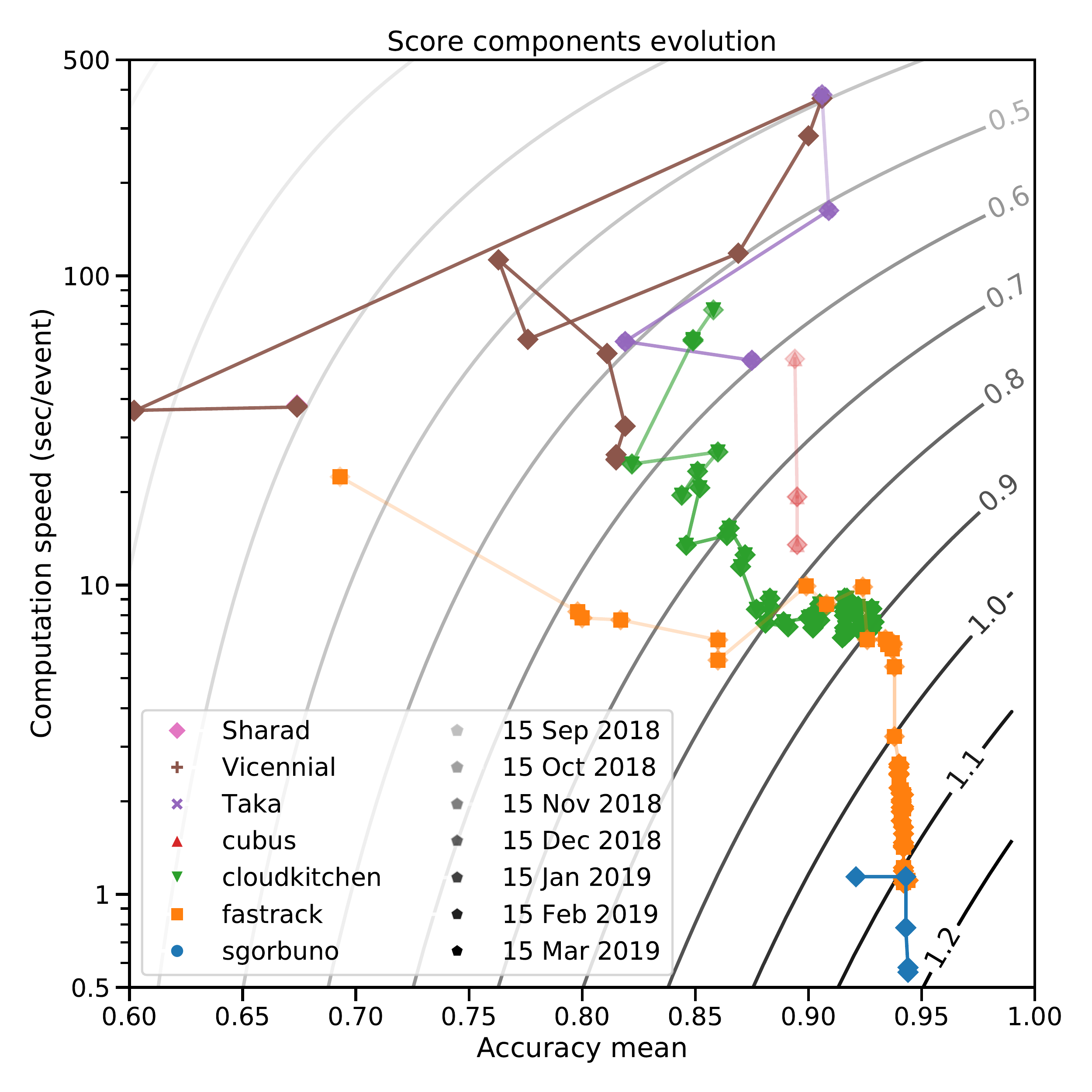}
\caption{(from\cite{TrackMLThroughput2020})
TrackML Throughput phase participants score evolution. 
The horizontal axis is the mean accuracy over the 50 test events, and the vertical axis is the average computation speed per event.
The total score, function of both variables, is displayed in grey contours.
Each colour/marker type corresponds to a contributor, the lines help to follow the score evolution.
}
\label{fig:ScoreEvolComp}
\end{figure}

For the Throughput competition, the simulation has been slightly adjusted. An ad-hoc score combining the Accuracy and the speed has been devised. The iso-score lines appear Fig.\ref{fig:ScoreEvolComp}, pushing participants to arrive closest to the bottom right corner, with the largest Accuracy score and smallest per-event time. 
Participants got a non zero score only if their submission could achieve more than 50\% accuracy (to avoid poor but super-fast algorithm) in less than 600~s per event (to avoid straining the resources dedicated to the challenge).

The Throughput evaluation required a reliable measurement of the inference time on well-defined resources, which was not possible (at least at the time) on Kaggle. It was then chosen to have the Throughput phase on Codalab which offers this possibility. It meant developing the code which would run the submitted software in a Docker environment with resources limited to 2 CPU and 4GB of total memory, and the code evaluating the Accuracy score (for the Accuracy phase this was done by the Kaggle Data Scientist relying for a large part on existing internal Kaggle code).
 The participant code is embedded in a skeleton taking care in particular of the reading of the event data and writing out the solution so that the time measured is purely the one of the inference. Several issues were uncovered and solved: 
 \begin{itemize}
     \item the time measurement was found to be reproducible only within 2\%, which could have lead to a change of ranks in case of many participants. Hence it was decided to measure the time after the end of the competition by averaging multiple (10) runs on a new dataset.
    This was done after the end of the competition and the updated time measurements were very close to the one provided online.
    \item it could have been the case that participants write in the log file useful information on the test dataset, and then use it in a later submission. To avoid this, logging was completely disabled. 
    \item it was expected (but not enforced) that participants would submit their source code which would be compiled on the platform. Uploading additional libraries was allowed given it was not expected the Docker environment to be complete with all possible utilities. However, some participants chose to upload their code directly as a library, which prevented the organisers to see their code during the competition.
    \item all sophisticated hacks could not be excluded; instead of multiple safety measures, for which the organisers had no time nor expertise, hacking was forbidden in the rules of the challenge, and the submission of the software required to win any price was expected to deter hacking effectively. No sign of hacking was detected after the competition.
  \end{itemize}

\subsection {Competitions} 

The TrackML Accuracy phase has run on Kaggle 1st May 2018 to 10th August 2018.
The TrackML Throughput competition opened a few weeks later, the 3\textsuperscript{rd}~September~2018. It was initially due to end 18\textsuperscript{th}~October~2018, but given the lack of competitors, it was extended till 15\textsuperscript{th}~March~2019.

\begin{figure}[ht]
\includegraphics[width=\textwidth]{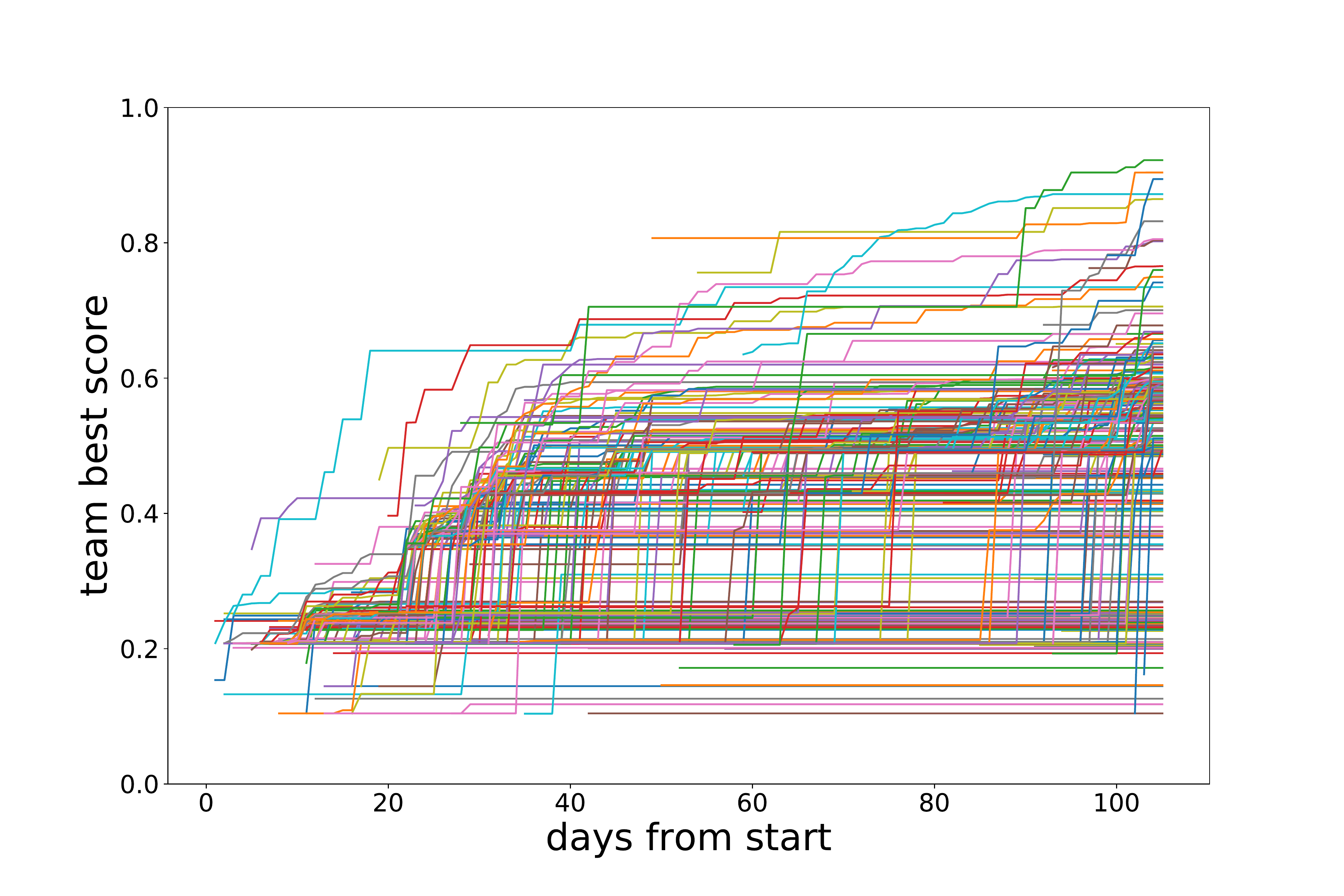}
\caption{(from \cite{TrackMLThroughput2020}) Evolution of the best score of each team as a function of time.}
\label{fig:AccuracyScoreEvo}       
\end{figure}

The Accuracy phase was well attended, with a total of 656 participants. Fig.\ref{fig:AccuracyScoreEvo} shows the evolution of the leader scores throughout the competition.  There is initially a large cluster of candidates achieving a score of 20\% to 25\%, which corresponds to the 22\% performance of the DBSCAN starting kit. After around 30 days, public kernels achieving a performance greater than 50\% (still based on DBSCAN) were posted on the public forum by some participants, which leads to a second group of candidates reaching a performance of 50\% to 60\% after 40 days of competition. Finally, a score of more than 90\% was only reached in the last days of the competition. Front runners are well separated from the pack and from each other, which is a clear indication of the complexity of the competition (if this had been a Tour de France stage, it would have been a mountain stage rather than a peloton finish in a flat stage). 

The competition forum has been very active with participants posting visualisation notebooks and algorithm kernels. The accompanying documentation provided minimal information on existing HEP tracking algorithms in order to not bias the competition towards existing solutions. However, participants know how to google and have searched and found and posted in the forum technical papers on tracking, courses and even the PhD dissertation of one of the organiser.

The post-competition analysis has uncovered that a variety of approaches was used, among which:

\begin{itemize}
    \item DBScan is a popular (in ML) clustering algorithm, to build clusters of nearby points in a space of n-dimension. The points belonging to a track are on an arc of helix, they are not close in 3D geometrical space. However, they can be brought to be close after suitable transformations. A simple example was provided to all participants as a starting kit, which allowed to give non-random results with a few lines of a code. Somewhat to the surprise of the organisers, the algorithm was further refined by many participants and all the way up to rank \#9 by \texttt{CPMP} who was given one jury prize.
    
    \item Hough transform, a classical (in HEP) algorithm mapping the geometric space to the track parameters space where the clustering is done was used by some participants and brought to rank \#7 by \texttt{Yuval and Trian} using several tricks to make it work in this situation.
    
    \item More innovative, the \texttt{finnies} have used Recurrent Neural Network to do the track following reaching rank \#12

    \item The most astonishing algorithm was from \texttt{outrunner} rank \#2 with a combination of a Neural Network and brute force: it first trains a NN to regress the probability that any pair of points belong to the same track. Then at the inference stage, it builds the large 100k$\times$100k matrix with the probability of all possible pairs of point. And finally, it builds the tracks by picking one by one the most likely pairs. It does work but is very compute-intensive, about one full day per event, which makes it unpractical.
    
    \item The post-challenge performance analysis has revealed somewhat accidentally that \texttt{diogo}, rank \#100 was the only one able to reach high efficiency for rare abnormal tracks coming far from the origin. This was achieved with an algorithm keeping track of connection between large voxels. It is unpractical as soon as there is some density of similar tracks but it is quite interesting for abnormal track finding.
        
    \item The classical (in HEP) track following algorithm has been improved with non-classical techniques by top performers, in particular \texttt{Top Quarks} rank \#1, \texttt{Sergey Gorbunov} rank \#3, \texttt{demelian} rank \#4, they will play an important role in the Throughput phase detailed later.

\end{itemize}

It was hoped that many participants (or new ones) will carry on in the Throughput phase. This did not happen, despite the extension of the deadline from October 2018 to March 2019. Only seven participants got a non zero score.
In hindsight, this can be understood to come from a combination of factors:
\begin{itemize}
    \item the lesser popularity of Codalab compared to Kaggle, where people can earn points across competitions.
    \item the complexity of the problem.
    \item the necessity to write C++ code, when a typical Kaggle participant is more used to python
    \item given the threshold at less than 600~s per event and more than 50\% efficiency, it was already non-trivial to have a non-zero score. 
    \item despite all the efforts to document and streamline the procedure to submit a solution, it still required a larger commitment than for a typical Kaggle competition. Also, the fact that log files were hidden to participants made debugging more difficult for them.
\end{itemize}

Nevertheless, the small number of participants has been more than compensated by the high quality of the top three participants (see Fig.\ref{fig:ScoreEvolComp}), who have all reached above 90\% accuracy with a time up to 0.5~s, while the original goal was around 10~s per event.

 The original idea was that algorithms developed in the Accuracy phase would be optimised and adapted to the second phase, possibly not by the same participants. This was not enforced in any way but in fact, it happened:
\begin{itemize}
\item Sergey Gorbunov (pseudo \texttt{sgorbunov})  rank 1 in Throughput phase had obtained rank 3 in the Accuracy phase 
\item Dmitry Emliyanov (pseudo \texttt{fastrack}) rank 2
in the Throughput phase had obtained rank 4 in the Accuracy phase (with pseudo \texttt{demelian})
\item Marcel Kunze (pseudo \texttt{cloudkitchen} rank 3 in Throughput phase ) used as a starting point the algorithm of \texttt{TopQuark}, rank 1 in the Accuracy phase, and has largely augmented it
\end{itemize}

\subsection {Scientific Conclusion}

 It is not possible to compare directly to in-house algorithms which would need to be adapted to this specific dataset. Also, they usually ignore the numerous tracks with $p_T$ less than 400~MeV (the tracks with the highest curvature) while algorithms presented here can reconstruct tracks down to 150~MeV. It can be estimated that in-house algorithms are not faster than 10~s per event on one CPU core, so one order of magnitude slower than Mikado from Sergey Gorbunov (a.k.a \texttt{sgorbuno}), 0.5s on two CPU cores. On the other hand, several simplifications were done in the dataset (in particular neglecting the sharing of points between tracks) so that it remains to be seen whether the new algorithms can live up to expectations when used in the ATLAS and CMS experiment context. The community is now in the process of doing this exercise.

In the end, Machine Learning was not at the core of the three best Throughput algorithms. Nevertheless, after extended discussions between the three winners and experts in the field, a consensus appears that there are two likely avenues for the use of Machine Learning in such problems (i) combine ML with classical discrete optimisation, for example using a classifier to select early and quickly the best candidates as done by Marcel Kunze a.k.a \texttt{cloudkitchen} (ii) use ML to automatise the lengthy tuning of the internal parameters of the algorithms (circa 10.000 in the case of Mikado by Sergey Gorbunov a.k.a \texttt{sgorbuno}).

\subsection{Organisation Conclusion}

The organisation of the TrackML challenge was a long process, the main elements of the timeline are indicated below:
\begin{itemize}
    \item March 2015 Berkeley Initial discussion Connecting The Dots workshop\footnote{\url{https://indico.physics.lbl.gov/event/149/}}
    \item March 2016 Vienna More discussion Connecting the Dots workshop\footnote{\url{https://indico.hephy.oeaw.ac.at/event/86/}}, team is being set up
    \item April 2017 2D hackathon Orsay Connecting The Dots workshop\footnote{\url{https://ctdwit2017.lal.in2p3.fr}}. Follow up paper released end 2017\cite{TrackMLRamp2017}.
    \item May 2017 first contacts with Kaggle
    \item March 2018 Connecting The Dots workshop Seattle\footnote{\url{https://indico.cern.ch/event/658267/}} 3D hackathon, feedback on almost final dataset and score
    \item May-August 2018 Accuracy challenge on Kaggle\footnote{\url{https://www.kaggle.com/c/trackml-particle-identification}}, follow-up paper released early 2019\cite{TrackMLAccuracy2019}.
    \item Dec 2018: NeurIPS competition workshop, with top participants invited
    \item Oct 2018-Mar 2019 : Throughput challenge on Codalab\footnote{\url{https://competitions.codalab.org/competitions/20112}}
    \item July 2019: CERN Grand finale workshop\footnote{\url{https://indico.cern.ch/event/813759/}} with top participants invited
    \item October 2019: Université Paris-Saclay Institut Pascal Advanced Pattern Recognition workshop\footnote{\url{https://indico.cern.ch/event/847626}} with top participants invited for two weeks
    \item End 2020 : final paper release \cite{TrackMLThroughput2020}
\end{itemize}
Also, there were more than 40 presentations at physics conferences (ICHEP, CHEP, EPS etc..), Machine Learning conferences (WCCI, NeurIPS CiML and Competition workshops), seminars in physics departments, python meetup (Paris, Gen\`eve) by all members of the team.

As can be seen, as the challenge develops, several events were organised to get feedback from the wide tracking experts community. These were important to exercise and adjust the mechanics of the challenge and discuss the conclusion and long term impact.

In particular, after the first round of initial discussions, a prototype has been the organisation of a challenge\cite{TrackMLRamp2017} on the RAMP\footnote{\url{https://paris-saclay-cds.github.io/ramp-docs/}} platform during the Connecting The Dots workshop\footnote{\url{https://ctdwit2017.lal.in2p3.fr}} (a workshop for experts in pattern recognition) held at IJCLab in Orsay in March~2017. 
The problem was essentially the same as the one exposed here but very much simplified to be a 2D problem with just 20~tracks per event (instead of 10.000 in 3D). There was no speed constraint.  The same accuracy score was used for the first time. This 2D challenge has already yielded a variety of algorithms (not directly applicable in 3D though) and demonstrated that the accuracy score was indeed selecting the best algorithms. Its success set a green light to launch the full project.

The team comprised 19 people, senior scientists with expertise in the field of tracking or Machine Learning, post-docs and students, all of them part-time. The total effort can be estimated to be 3 Full-Time Equivalent year. 
The main tasks were: preparing the dataset, the accompanying documentation, helper library and starter kit for the 4 hackathons and competitions organised, interacting with Kaggle, implementing the competition in Codalab, searching for sponsors, running the competition, organising the different associated workshops, doing the post-competition analysis and writing the papers. 

Sponsoring was needed for the prizes (30~k\$ and 1~NVidia V100 for the Accuracy phase, 15~k euros and 1 NVidia V100 for the Throughput phase) and for the invitations to NeurIPS 2018 Competition workshop (Accuracy phase) and CERN July 2019 (Throughput workshop).

\subsection{Follow-up studies}

Separately, the availability of the TrackML dataset has been extremely useful to facilitate the collaboration of experts which are usually working on their own data within their own experimental team.
It has been used for new studies like investigating tracking with simulated annealing on a D-Wave quantum computer\cite{Zlokapa:2019tkn}, or with graph networks\cite{Ju:2020xty,Duarte:2020ngm}.
Somewhat unexpectedly, the dataset has also been used to explore the usage of Augmented Reality to visualise scientific data\cite{Wang2020,wangphd}. 
The dataset will be released in a near future on CERN open data portal. 

\section{LHC Olympics}
\label{s:lhco}
\subsection{Intro}
Despite an impressive and extensive effort by the Large Hadron Collider (LHC) collaborations, there is currently no convincing evidence for new particles produced in high-energy collisions. LHC Olympics 2020 Anomaly Detection Challenge challenge was aimed at exploring the capabilities of machine learning to enhance the potential signal of Beyond Standard Models (BSM) using all of the available information.

\subsubsection{The challenge goal}
The challenge goal was to ensure that the LHC search program is sufficiently well-rounded to capture “all” rare and complex signals. Different stages of the competition are focused on different volumes of the phase space since potential BSM parameter space is vast. 

\subsubsection{Challenge setup}
The LHC Olympics 2020 setup is aligned with the first LHC Olympics organized in 2005-2006\footnote{\url{https://public-archive.web.cern.ch/en/Spotlight/SpotlightOlympics-en.html} and \url{https://www.kitp.ucsb.edu/activities/lhco-c06}}. Participants are provided with two types of data:
\begin{itemize}
    \item “Monte Carlo Simulation Background”: This is a simulated sample that does not have a signal. Be warned that both the physics and the detector modelling for this simulation may not exactly reflect the “Data”.
    \item “Data”: These samples contain a mixture of background with some new signal(s). Three unique samples referred-to as \textit{black boxes} have been released during LHCO 2020 challenge. All the samples had become available in November 2019. The first sample has been unveiled mid-January, during winter part of the LHC Olympics, and the remaining two has been unveiled during the LHC Olympics summer workshop.
\end{itemize}
Both the “Simulation” and “Data” have the same event selection criteria (see Section \ref{sec:LHCO_data}). Participants had to find signals of BSM in “Data” samples and to report various metrics that estimate the confidence of those findings. There were two workshops during 2020 focused on the discussion of multiple techniques and intermediate challenge results. The organizational committee of the LHC Olympics 2020 coincides with the one of those workshops\footnote{Gregor Kasieczka, Benjamin Nachman and David Shih}. 

\subsection{Data description}
\label{sec:LHCO_data}
For both background and black box data, events supposed to have the form of X $\to$ hadrons, where X is a new massive particle with an $\mathcal{O}(\mathrm{TeV})$ mass. Events are selected after a single trigger of anti-$k_t$ $R = 1.0$ jet\cite{Cacciari_2008} with pseudorapidity $|\eta| < 2.5$ and the transverse momentum $p_t > 1.2$~TeV. Number of events per data sample is the same and equals to~1M. 

These events are stored as pandas DataFrames saved to compressed HDF5 format. For each event, all reconstructed particles are assumed to be massless and are recorded in detector coordinates ($p_t, \eta, \phi$). More detailed information, such as particle charge or type, is not included. Events are zero-padded to constant size arrays of 700 particles. The array format is therefore \texttt{(1M, 2100)}.

\subsubsection{Background}
The background sample of 1M events consists of QCD dijet events simulated using Pythia8 and Delphes 3.4.1. Both the physics and the detector modelling for this simulation are not guaranteed to precisely reflect the signal ``data". 

\subsubsection{Signal}
\begin{figure}
     \centering
     \begin{subfigure}[b]{.45\textwidth}
         \centering
         \includegraphics[width=\textwidth]{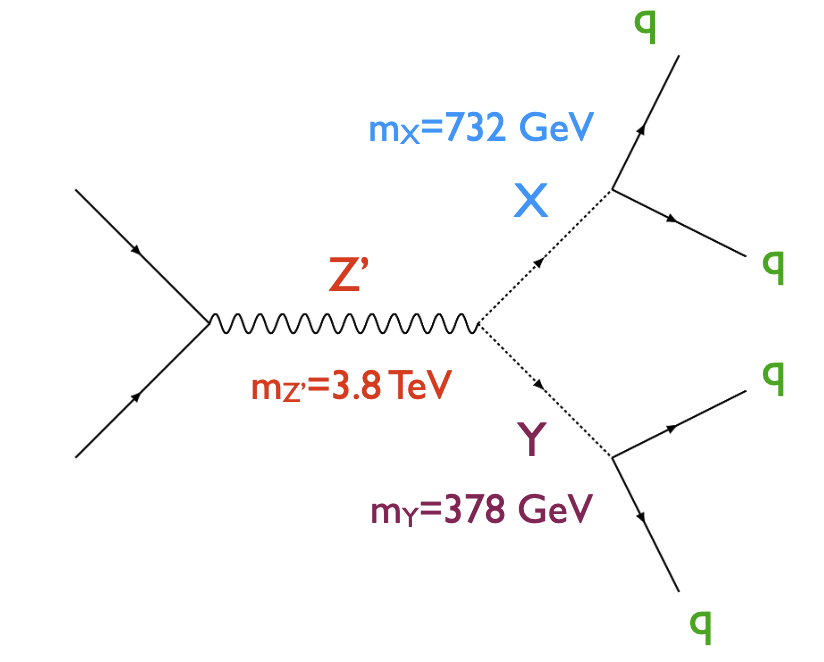}
         \caption{$Box1$}
         \label{fig:LHCO_box1}
     \end{subfigure}
     \hfill
     \begin{subfigure}[b]{.45\textwidth}
         \centering
         \includegraphics[width=\textwidth]{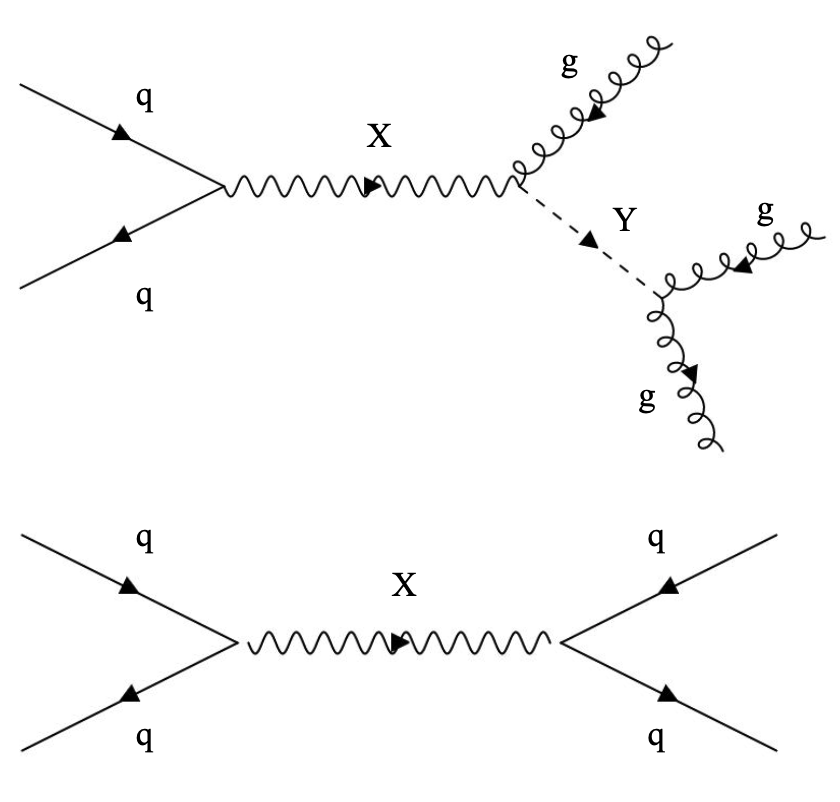}
         \caption{$Box3$}
         \label{fig:LHCO_box3}
     \end{subfigure}
        \caption{The new-physics modes hidden in the black boxes.}
        \label{fig:Two boxes}
\end{figure}

The signal dataset is split into three files, referred to \textit{black boxes}. Each ``black box" contains 1M events meant to be representative of actual LHC data. These events may include BSM signal(s), i.e. it contains either mixture of some signal and background or just background. The signal in the former case represents a kind of new physics simulated by the same software packages as the background. Fig.\ref{fig:LHCO_box1} represents the process hidden in the first box. There were 834 events of this kind. The second black box didn't contain any signal and was filled with the same QCD background events to check the participant's algorithm false positive rate at the boundaries of the phase space. The most complicated case was hidden in the third box. It required to stack together two decay modes depicted at Fig.\ref{fig:LHCO_box3} with $m_X=4.2$~TeV, $m_Y=2.2$~TeV, $BR(X \to qq) = 1-BR(X \to Y_g)=0.375$. Work\cite{Agashe_2017} inspires this physics, i.e., simple extensions of RS motivated by LHC Run I null results and little hierarchy problem, where X represents Kaluza-Klein gluon and Y - IR radion. The total number of signals in the third box was equal to 3200. 

The competition datasets are published at Zenodo archive\cite{kasieczka_gregor_2019_3596919}.

\subsection{Evaluation procedure}
Participants should report:
\begin{itemize}
    \item a p-value associated with the dataset having no new particles (null hypothesis);
    \item a description of the new physics, as complete as possible. For example the masses and decay modes of all new particles (and uncertainties on those parameters);
    \item number of signal events (+uncertainty) in the dataset (before any selection criteria).
\end{itemize}
Outcomes will be judged based on a) the optimality of the p-values and b) the accuracy of the new physics characterization:
\begin{itemize}
    \item optimality corresponds to the ``best" p-value will be the lowest reported p-value that is above the fully optimal p-value (as determined with a fully supervised deep learning classifier);
    \item accuracy is computed by the number of sigmas from the right answer is used wherever applicable. Number of sigmas is estimated as $|(\mathrm{predicted} - \mathrm{true}) / \mathrm{predicted\_uncertainty}|$
\end{itemize} 
Organizers prepared the competition starting kit\footnote{\url{https://github.com/lhcolympics2020/parsingscripts}} with scripts that read in the data and perform exploratory data analysis with it. 

\subsection{Prizes and participation statistics}
\label{sec:LHCO_stats}
There was no money prize associated with the competition and, perhaps, it was mainly meant for particle physicists since the entry required some understanding of basic QCD models and new physics models. Thus, it has attracted a couple of dozens of participants during 2020 mostly from physics departments. It was also relatively lightweight in terms of evaluation tools; competitors had to submit all the metrics via a google form. Such form allowed to collected extended feedback like a description of the new physics a participant was aiming for. The organizers put together two workshops during the winter\footnote{\url{https://indico.cern.ch/event/809820}} and the summer\footnote{\url{https://indico.desy.de/event/25341}} of 2020. Every workshop has many relevant contributions to the inclusive search for the new physics. Detailed workshop outcome analysis is available as a workshop contributions\footnote{\url{https://indico.cern.ch/event/809820/contributions/3708303/attachments/1971116/3347225/SummaryTalk.pdf}, \url{https://indico.desy.de/event/25341/contributions/56822/attachments/36777/45997/SummaryAnomalyDetectionWorkshopJuly2020.pdf}}. 

\subsection{Conclusion}
The LHC Olympics presents a charming and successful format for running a competition in sustainable and cost-saving mode. It is more focused on the physic-oriented results rather than attracting a broader data-science audience. The main page of the challenge\footnote{\url{https://lhco2020.github.io/homepage/}} includes references to several papers describing participant's contributions. Also, there is a community whitepaper on the competition outcomes scheduled to be published. It welcomes every participant for the co-authorship. LHC Olympics organizers invite new BSM black boxes from the community for the future runs of the challenge.

\section{Competitions platforms}
\label{s:platforms}


\subsection{Platforms for data challenges}
There is a dozen of platforms that allow hosting data challenges. These differ significantly in terms of flexibility, functionality and community factors. As it was mentioned in the previous sections, challenges requirements can be quite diverse and demanding. This section gives an overview of the leading players and alternative approaches that can be used to run a new challenge. A platform for hosting a data challenge is a service that is provided by a company or institution behind it. Those services usually follow the so-called Common Task Framework (CTF)(see Sec.\ref{intro:CTF} and paper~\cite{donoho50years} for details).

Historically different groups started to develop such services around different challenges; thus, its functionality may differ. Anyway, since those follow the common competition protocol, every platform allows to upload a dataset, describe challenge condition, setup evaluation procedure and invite the community to a new challenge. However, some platforms are better for dealing with human-in-the-loop evaluations; some give better flexibility in terms of metric specification; some can deal with private data. We are going to overview the features of the following platforms:

\begin{itemize}
    \item AICrowd\footnote{\url{https://www.aicrowd.com/}} by EPFL, AICrowd, 
    \item CodaLab\footnote{\url{https://competitions.codalab.org/}} by Universit\'e Paris-Saclay,
    \item CrowdAnalytiX\footnote{\url{https://www.crowdanalytix.com/}} by CrowdAnalytiX,
    \item EvalAI\footnote{\url{https://eval.ai/}} by CloudCV,
    \item Kaggle\footnote{\url{https://www.kaggle.com/competitions}} by Alphabet Inc,
    \item RAMP\footnote{\url{https://ramp.studio/}} by Paris-Saclay Center for Data Science,
    \item Tianchi\footnote{\url{https://tianchi.aliyun.com/competition/}} by Alibaba,
    \item TopCoder\footnote{\url{https://www.topcoder.com/challenges?tracks[DS]=true}} by TopCoder.
\end{itemize}
The list is not meant to be comprehensive, as it is focused on active platforms with broad communities as of the end of 2020, i.e. with more than ten competitions started during 2020 using publicly available information. Also, there are platforms like Grand Challenge\footnote{\url{https://grand-challenge.org/}} that are focused on some narrow scientific domain. We have identified the following criteria for comparing the platforms above that are relevant for HEP-related competitions. 

\subsubsection{Criteria}
We outline the main characteristics that we use for the comparison. Those criteria are sorted by order of relevance to the competitions described in this book chapter.

\textbf{Code sharing, reproducibility:} challenge participants are not always motivated by getting the highest score. Instead, they might want to explore new things or to get praised by the community. Thus, the ability to share and discuss their code with other participants becomes a crucial feature for new complicated challenges like the HEP ones. Sometimes code-sharing is available right within the platform, like at Kaggle, or sometimes participants can link their solution to github repository/commit, so other participants can reproduce and play with it. Such feature adds greatly to the reproducibility of the winning solutions making it much more scientific.

\textbf{Code submission:} accuracy metrics are not enough to evaluate the dynamic aspects of a participant's code. In some cases like tracking or triggering, one may be interested in comparing the accuracy of an algorithm only if certain execution speed/resource consumption constraints are met. Thus, some platforms support code as a submission to evaluate a solution to the full extent.

\textbf{Community activity:} it is a cumulative estimation that takes into account the total number of challenges organized, number of challenges in 2020, the maximum number of participants per challenge and estimated size of the community. 

\textbf{Custom metrics:} the ability to implement custom metrics is crucial for some non-trivial cases that wish to compare algorithm performance for non-usual challenges like it was for Flavour of Physics. Sometimes it is needed to make a trade-off between accuracy and performance like it was for TrackML. Some platforms allow choosing just one among many predefined metrics; some allow for custom implementations. Some platforms charge an additional cost for implementing non-standard metric. 

\textbf{Staged challenges:} sometimes challenges might look too weird at the beginning, so it helps to split it in smaller chunks. Thus, it is possible to mitigate risks of data leakage by adding a preliminary stage and testing the competition settings. It will help people to keep the context between stages and smoothly transfer knowledge of the best solutions.

\textbf{Private challenge evaluation:} data privacy is a serious issue even in fundamental science, so some platforms allow running participant's solutions evaluation using an organizer's dedicated machines, thus one setup a challenge without the need to share restricted datasets.

\textbf{Open-source:} in usual scenarios, a platform operates as a service and challenge organizers do not care much about tweaking its functionality. However, open-source gives the ability to evaluate the project activity, check the details of platform evaluation mechanics or run own instance of the service for some local events with private datasets, for example.

\textbf{Human-evaluation:} some challenges do not have ground-truth labels in the data. For example, evaluation of a dialogue bot requires communication with a living person, or images of galaxies labelled by an agent may require extra human validation. Some platforms allow connecting to human-evaluation platforms such as Amazon Mechanical Turk (see below) or alike.

\textbf{Reinforcement-Learning (RL) evaluation:} agents designed to operate in a dedicated environment present another challenging task for a fair evaluation for a couple of reasons: a) environments can be very diverse, b) each agent may require considerable computational resources the platform needs to account for, c) each agent operates in randomized environments thus it may require several evaluations to get a statistically-sound score. 

\textbf{Run for free:} many platforms allow setting up money prizes to the competition winners, so they charge for running those settings. Some platforms still allow to run a challenge almost for free - if the problem is not computationally heavy, it is possible to run it using the service infrastructure. Sometimes it is possible to connect the organizer's computational resources to the service, thus avoiding extra charges. 

\subsection{Comparison}
Overview of the platforms concerning the criteria above is presented in the Tab.~\ref{tab:platforms}. In addition to the overall comparison, it is worth mentioning individual features that are difficult to fit into a generic table. 

\begin{table}
\tbl{Platform overview.}{%
\begin{tabular}{l c @{\hspace{0.7ex}} c @{\hspace{0.7ex}} c @{\hspace{0.7ex}} c @{\hspace{0.7ex}} c @{\hspace{0.7ex}} c @{\hspace{0.7ex}} c @{\hspace{0.7ex}} c }
\hline Criteria & AICrowd	&	CodaLab	&	CrowdAnalytiX	&	EvalAI	&	Kaggle	&	RAMP    &   Tianchi	&	TopCoder \\
\hline Code-sharing & \xmark & \cmark & \xmark & \cmark & \cmark & \cmark & \cmark & \xmark\\
\hline Code submission & \cmark & \cmark & \xmark & \cmark & \cmark & \cmark & \cmark & \cmark\\
\hline Active community & \dstarr & \dstarrr & \dstar & \dstarr & \dstarr\dstarr &  \dstar  & \dstarrr & \dstarr\\
\hline Custom metrics & \cmark & \cmark & \cmark & \cmark & \cmark & \cmark & ? & \xmark\\
\hline Staged challenge & \cmark & \cmark & \xmark & \cmark & \xmark & \cmark & \xmark & \xmark\\
\hline Private evaluation & \xmark & \cmark & \xmark & \cmark & \xmark & \xmark & \xmark & \xmark\\
\hline Open-source & \cmark & \cmark & \xmark & \cmark & \xmark & \cmark & \xmark & \xmark\\
\hline Human evaluation & \xmark & \xmark & \xmark & \cmark & \xmark & \xmark & \xmark & \xmark\\
\hline RL-friendly & \cmark & \xmark & \xmark & \cmark & \xmark & \xmark & \xmark & \xmark\\
\hline Run for free & \xmark & \cmark & \xmark & \cmark & \cmark & \xmark & ? & \xmark\\
\hline
\end{tabular}}
\label{tab:platforms}
\end{table}

\textbf{RAMP:} Rapid Analytics \& Model Prototyping is a service that is mainly used by the Paris-Saclay Center for Data Science to support own events like hackathons or datacamps. Remarkably, RAMP involves two phases of each event - competition and collaboration. During competition phase participants, try to design their algorithms, while upon collaboration stage, they share their approaches and team up for the sake of a better solution. RAMP is published under BSD-3 license\footnote{\url{https://github.com/paris-saclay-cds/ramp-board/}}, so it may come handy for a lightweight setup of an event at own premises. 

\textbf{Kaggle:} allows to run a competition entirely free for non-commercial purposes in so-called ``InClass" mode: a) so Kaggle does not advertise it to the community and b) gives a limited setup flexibility, i.e. one cannot evaluate submissions against anything but the set of pre-defined straightforward metrics like RMSE or ROC AUC. Also, such competitions do not award any Kaggle ranking points to the participants, which reduces the incentive to join it significantly. Once a company/university decides to run a full-fledged public challenge, it is possible to implement a custom metric, but it may turn out to be quite expensive.

\textbf{Tianchi:} despite a relatively young age, it is a top-rated service in China with very similar to Kaggle functionality that includes running kernels and ranking points. Challenges quickly can gain several thousand participants. However, most of the audience is Chinese, so communication skills in Chinese would come handy.

\textbf{TopCoder:} is one of the oldest and biggest worldwide platforms for outsourcing coding tasks. Thus the audience is huge - 1.5 million of users. However, it added machine learning tasks in 2018, so the only fraction of the total users is relevant for addressing data-driven challenges.

\subsection{Alternative approaches}
The platforms from the comparison above implement challenges along Common Task Framework\cite{donoho50years}. However, it is not the only option. Below is a list of platforms that rely on different assumptions and implement peculiar interaction protocols. 

\textbf{Amazon Mechanical Turk (AMT)\footnote{\url{https://www.mturk.com/}}:}  is a marketplace for completion of virtual tasks that require human intelligence. A business or academics typically use it to label data that later on can be used for training ML algorithms. AMT has been around for more than 15 years. Major companies like Google and Microsoft have similar versions of such marketplaces.

\textbf{Zooniverse\footnote{\url{https://www.zooniverse.org/}}:} While AMT focuses on pretty generic tasks like reading labels from images, captcha translation, listening comprehension, tagging inappropriate images, etc. Zooniverse builds a community of people that are interested in contributing their efforts and intelligence to scientific research advances. It provides participants with unlabelled datasets from a wide variety of scientific branches: biology, climate, history, physics, etc. Those datasets require human intelligence not only for labelling but also for understanding the scientific assumptions of the domain and phenomena presented. Participation in real-science research can motivate people quite significantly. There are cases when discussions between scientists and Zooniverse participants lead to new scientific discoveries\cite{clery2011galaxy}. 

\textbf{OpenML\footnote{\url{https://www.openml.org/}}:} is an online machine learning platform for sharing and organizing data, machine learning algorithms and experiments. Founders of the platform are passionate about the comparison of different ML methods. Thus they have created the service that allows to run an algorithm across different datasets and systematically compare its performance. While there are no private leaderboards, every check is performed via system API and protocol systematically. Thus new experiments are immediately compared to state of the art without always having to rerun other people’s experiments. The recent development of OpenML involves the design of AutoML evaluation framework for a broad spectrum of datasets.

\textbf{PapersWithCode (PwC)\footnote{\url{https://paperswithcode.com/}}:}  organizes access to technical papers that also provide the software used to create the paper’s findings, has grown immensely in the past few years. With the help of this platform, one can find the most current state of the art to the problem of the interest and read details of the method in the linked paper from arXiv.

\textbf{InnoCentive\footnote{\url{https://innocentive.wazoku.com/}}:} is an innovative hub for a new kind of problem-solving. It describes the framework of ‘Challenge Driven Innovation’ (CDI) that helps to reformulate a task at hand into a series of modules or challenges that are addressed later either by a network of so-called \textit{solvers} or internal company members. CDI have examples of different kind of challenges, including idea, validation, proof of concept, prototype, and production. So it is not specific to data labelling or algorithm development. 

\textbf{Seasonal events:} there are many yearly data analysis events organized around the world. Usually, those are hosted by universities and attract quite a significant number of participants. International Data Analysis Olympiad (IDAO)\footnote{\url{https://idao.world}} is just a single example among many\footnote{Data Mining Cup, \url{https://www.data-mining-cup.com/}},\footnote{ASEAN Data Science Explorers\url{https://www.aseandse.org/}}. IDAO has engaged more than 2500 participants across 83 countries in 2020. Interesting and unique challenges might fit such events very well, and in that case, organizers will alleviate the burden of preparing and running the challenge quite significantly.

\textbf{Other:} There are many different venues for interactions between science and citizens. Michael Nielsen gives a good overview in his book ``Reinventing Discovery: The New Era of Networked Science"\cite{nielsen2020reinventing}. A remarkable example of such interaction is the design of a network of micro-prediction agents that follow specific question-answering protocol. Authors of those agents get rewards for providing correct answers. Such protocol gives incentive to the participants to come up with better algorithms and suitable external data sources\cite{cotton2019self}. A broader list of citizen-science projects is, of course, available at Wikipedia\cite{wiki_crowdsourcing}. 



\section {Open datasets and repositories}
Several datasets prepared for challenges are listed in the following.
In addition, the LHC experiments have released some fraction of their data with corresponding simulated events (with ground truth) but there are no associated metrics. Also, authors of papers on the application of Machine Learning techniques to High Energy Physics are often willing to share their datasets on request, even if not formally released. 

\begin{itemize}
    \item HiggsML dataset is available on the CERN Open Data portal~\cite{higgsml_cern_odp_dataset} with accompanying documentation~\cite{higgsml_cern_odp_document}. All 818,238 events have been released including ground truth, while only a subset of 250,000 events is available on Kaggle\footnote{\url{https://www.kaggle.com/c/higgs-boson}}. 
    For each event, it lists 17 low-level features and 13 high-level features for two classes. Beyond classification, a python script allows to introduce systematic effects \cite{victor_estrade_2018_1887847,estradephd}.
    \item Flavour of Physics challenge dataset\footnote{\url{https://www.kaggle.com/c/flavours-of-physics/data}}.
    \item Datasets for the TrackML challenge : (i) the Kaggle one used for the Accuracy phase \footnote{\url{https://www.kaggle.com/c/trackml-particle-identification}} 
    (ii) the Codalab one used for the Throughput phase\footnote{\url{https://competitions.codalab.org/competitions/20112}}. 
    Compared to the Accuracy one, a few features were corrected 
    (iii) the CERN Open Data Portal final release in preparation 
    \item LHC Olympics-2020 dataset\cite{kasieczka_gregor_2019_3596919}.
    \item LHCb Muon Identification challenge dataset \footnote{\url{https://www.kaggle.com/kazeev/idao2019muonid}} was published within International Data Analysis Olympiad-2019\footnote{\url{https://idao.world/history/\#idao-2019}}.
    \item LHCb PID compression challenge\footnote{\url{https://zenodo.org/record/1231531}}, with baseline solution\footnote{\url{https://github.com/weissercn/LHCb_PID_Compression}}.
    \item MiniBooNE Particle Identification dataset\footnote{\url{https://www.kaggle.com/ukveteran/miniboone-particle-identification}}.
    \item LArTPC 2D/3D Simulation for Particle Segmentation \& Clustering\footnote{\url{https://osf.io/vruzp/}}.
    \item Particle Identification from Detector Responses, a simplified dataset of a GEANT based simulation of electron-proton inelastic scattering measured by a particle detector system\footnote{\url{https://www.kaggle.com/naharrison/particle-identification-from-detector-responses}}.
    \item the top tagger dataset\cite{Top_tagger_dataset} has been used for extensive studies of top quark tagging\cite{Kasieczka:2019dbj}. 
\end{itemize}

There are several catalogues that reference HEP-related dataset, which can be handy for adding a published dataset to increase its visibility:

\begin{itemize}
    \item CERN Open Data Portal\footnote{\url{https://opendata.cern.ch}} hosts a collection of datasets from all large LHC experiments as well as from OPERA experiment.
    \item The Durham High-Energy Physics Database\footnote{\url{https://www.hepdata.net/}}. It hosts the data points from plots and tables related to several thousand publications including those from the LHC collaborations. It does not hold any  event datasets unlike the CERN ODP.
    \item UCI ML HEP portal\footnote{\url{http://mlphysics.ics.uci.edu}} hosts a variety of HEP datasets associated with published papers, in particular the HIGGS UCI dataset \cite{HIGGS_UCI_dataset} produced for the study \cite{Baldi:2014kfa}.  
    \item Inter-Experimental LHC Machine Learning (IML) Working Group datasets\footnote{\url{https://iml.web.cern.ch/public-datasets}}.

\end{itemize}

\section {Guidelines for new competition organisers}

As for a movie or a novel, there are no rules which would guarantee the success of a scientific challenge. However, a set of guidelines can be derived from the experience gathered from the challenges summarised in this chapter, which does not pretend to be exhaustive. 

The overarching goal of a Machine Learning challenge is the scientific issue. It should be compelling both for experts of the domain (High Energy Physicists), for experts in Machine Learning, and non-experts. It should be possible to pitch it to someone with  little scientific background. At the same time, it should appear complex enough to be interesting, and non-physicists should feel they can contribute without a big disadvantage compared to physicists.
It is important to focus on just one issue. For example, for an event classification problem like HiggsML or Flavour or Physics, one would provide particle 4-vectors and hide all the complexity of accurate calibration of the detector.

The centrepiece of a challenge is the dataset. It will have a life well beyond that of the challenge. In some sense, a challenge can be seen just as a way to advertise a dataset. The dataset should be curated and prepared to be easily understood and handled by non-experts, preferably with no need for non-standard tools. It should still retain some richness has the same dataset can be used later on for other challenges, tutorials, benchmarks. The preparation of the dataset is probably the more time-consuming part of a challenge preparation.

Since a challenge is by nature a competition, there should be a unique score to rank the participants.   Domain experts are not used to ranking techniques based on a single number as they would typically like to see in-depth studies (with many curves and histograms) concerning various merit of a technique. Yet, there should be a single score, defined before the competition. Participants will optimize for this score, and the organisers bet that at the end of the competition the best algorithms from the point of view of the score, will also be the best algorithms for domain experts. Besides, the score should be sufficiently simple to be understood by the non-experts and stimulate their creativity (not a black box), robust against possible ``hacks", and, with limited luck factor when used in a challenge context. For example, in the HiggsML competition, the AMS (Equation~\ref{eq:higgsml:score}) was chosen as it was much more relevant for a typical HEP classification problem (which are very unbalanced) than the usual ROC-AUC, and the regularisation term 10 allowed to reduce the statistical uncertainty on the evaluation. Defining the score is probably the most difficult part of a challenge preparation.

Running a prototype of the challenge as part of an expert workshop or grad student school allows debugging many issues, from misleading documentation to the mechanics of the challenge platform. 

Finally, challenges are a competitive market. Successful participants will spend months on a particular challenge, but they decide in little time in which competition they will enter. Without going into any details, this drives much of the effort in relying on an established challenge platform, on streamlining the challenge documentation (which should be readable by any scientific undergraduate and at the same time open up to more complex knowledge) and starting kit. In particular, submitting a first ``hello world" solution should be possible in less than an hour. Public Relations is also important, as well as foreseeing incentives for participation. Money incentives are good but their role should not be over-emphasized. Invitations to participate in workshops at Machine Learning conferences or major HEP laboratories like CERN are valued by participants. 

It can never be expected that the outcome of a challenge will be a piece of software ready to be plugged in. It is rather a smorgasbord of algorithms, well documented or not, forum posts or blogs.

Post-challenge workshops have the merit to keep some participants engaged in a collaboration with physicists (others will immediately move on to another competition). 
Special ``jury" prizes set aside for algorithms judged on their overall merit (not the absolute best score, but also novelty, usability...) allow keeping these participants engaged.
Offering them the possibility to contribute to post-challenge papers is another means. The post-challenge phase is particularly interesting when it allows real collaboration, combining several good ideas, compared to the competition phase which is, well, a competition: discussion on the forum happens, notebooks are exchanged, but the best competitors are often silent until the end.

\section {Conclusion}
\label{sec:conclusion}

In this chapter, four quite different High Energy Physics scientific competitions have been summarised. 
In all cases, new approaches have emerged, in addition to the optimisation of existing ones. 
In all cases, the formal end of the competition is actually the beginning of a new effort to sift through the wealth of information generated. Also, a long-lasting impact of the competitions is the dataset released with accompanying metric. 
High Energy Physics boasts diversity and complexity of data structure and a variety of scientific questions raising interest well beyond its perimeter. It offers a wide range of future competition topics. Hopefully, resources, services and guidelines outlined in this chapter will help to pave the way for the design and organisation of new fascinating challenges. 

\clearpage
  
\bibliographystyle{tepml}

\bibliography{hep-ml-resources,challenges2}


\end{document}